\documentclass[journal]{IEEEtran}
\usepackage{xcolor}
\ifCLASSINFOpdf
   \usepackage[pdftex]{graphicx}
\else
\fi
\usepackage{amsmath}

\begin{document}
%
\title{Context-Aware Self-Supervised Learning of Whole Slide Images}
%
%
%

\author{Milan Aryal and Nasim Yahyasoltani 
\thanks{ The authors are with the Department
of Computer Science,  Marquette University, Milwaukee, WI, USA (e-mail: \{Milan.aryal, nasim.yahyasoltani\}@marquette.edu).}
\thanks{Part of this work  was presented at the IEEE ICASSP Conference held in Singapore, during May 7-10, 2022.}}
\maketitle

\begin{abstract}

Presenting whole slide images (WSIs) as graph will enable a more efficient and accurate learning framework for cancer diagnosis.  Due to the fact that a single WSI consists of billions of pixels and there is a lack of vast annotated datasets required for computational pathology, the problem of learning from WSIs using typical deep learning approaches such as convolutional neural network (CNN) is challenging. Additionally, WSIs down-sampling may lead to the loss of data that is essential for cancer detection. A novel two-stage learning technique is presented in this work. Since context, such as topological features in the tumor surroundings, may hold important information for cancer grading and diagnosis, a graph representation capturing all dependencies among regions in the WSI is very intuitive.  Graph convolutional network (GCN) is deployed to include context from the tumor and adjacent tissues, and self-supervised learning is used to enhance training through unlabeled data. More specifically,  the entire slide is presented as a graph, where the nodes correspond to the patches from the WSI.  The proposed framework is then tested using WSIs from prostate and kidney cancers. To assess the performance improvement through self-supervised mechanism, the proposed context-aware model is tested with and without use of pre-trained self-supervised layer. The overall model is also compared with multi-instance learning (MIL) based  and other existing approaches. 

\end{abstract}


%
\IEEEpeerreviewmaketitle

\section{Introduction}
The pathologic assessment of a tissue sample or excision is used to diagnose cancer. The sample is cut into small layers before being put on a glass slide and inspected under a microscope to determine the diagnosis. The diagnosis and prognosis of cancer is done by pathologists using hematoxylin and eosin(H\&E) and other stains slide \cite{5299287}. The developments of digital scanners in recent years have enabled the digitization of glass slides into WSIs. The WSIs are multi-scale, multi-resolution digital images of the tissue slides. Preliminary diagnosis by machine, obviously, may help pathologists be more efficient in diagnosis and may assist to avoid certain mistakes. The usage of WSIs in computational pathology for diagnosis has recently risen dramatically.   


In this work we are looking at the classification of the grading of cancer using deep learning. More specifically, the  WSIs of prostate and kidney have been used for grading the cancer. After lung cancer, prostate cancer is the second leading cause of mortality among men \cite{cancer_stat}. On the basis of the grading of prostate tissue samples, prostate cancer can be diagnosed. The pathologist grades the samples using the Gleason grading system, which is subsequently translated into an ISUP score on a scale of 1 to 5.   

Kidney cancer falls into top 10 most common cancers for both male and female. The kidney cancer accounts for 5\% of the new cases among male and 3\% of the new cases among female \cite{cancer_stat}. Renal Cell Carcinoma (RCC), which accounts for 85\% of all occurrences of kidney cancer, is the most prevalent malignant tumor \cite{SHUCH201585}. Based on the difference in histology, molecular characteristics and response to therapy RCC can be further classified into subtypes \cite{Kidney}. The major subtypes of the RCC include "Clear cell", "Papillary cell" and "Chromophobe" which account for 75\%, 16\% and 7\% of all the RCC cases, respectively \cite{SHUCH201585}. In this work, we will look into grading of these subtypes of the kidney cancer.

Deep learning has been used successfully in the field of medical imaging \cite{7890445}, \cite{10.5555/3298239.3298476}, \cite{10.1007/978-3-319-24553-9_63}, \cite{10.1007/978-3-642-40763-5_51}. These accomplishments, which are largely the result of supervised learning, are based on a large amount of labeled data. The availability of annotated datasets in the medical domain is extremely limited. The lack of annotation in medical data is  mainly because the task requires an expert in the field to go through each data one by one manually to label them. This makes the labeling process more expensive and time consuming. The use of transfer learning \cite{7950523}, \cite{tl}, \cite{heker2020joint} in medical images is widely considered to overcome this issue in the dataset. To extract the meaningful representation of unlabeled data, unsupervised learning techniques might also be recommended. It is possible to employ self-supervised learning \cite{9156540}, a type of unsupervised learning to learn from unlabeled data and then apply it to subsequent task. The application of self-supervised learning in the medical field as a pretext task has recently increased due to the effectiveness of the technique in other computer vision applications \cite{Azizi_2021_ICCV}, \cite{NEURIPS2020_d2dc6368}. Self-supervised learning approaches benefit from not requiring lots of labeled data. 

\begin{figure*}[htp]
    \centering
    \includegraphics[width=\textwidth, height = 10cm]{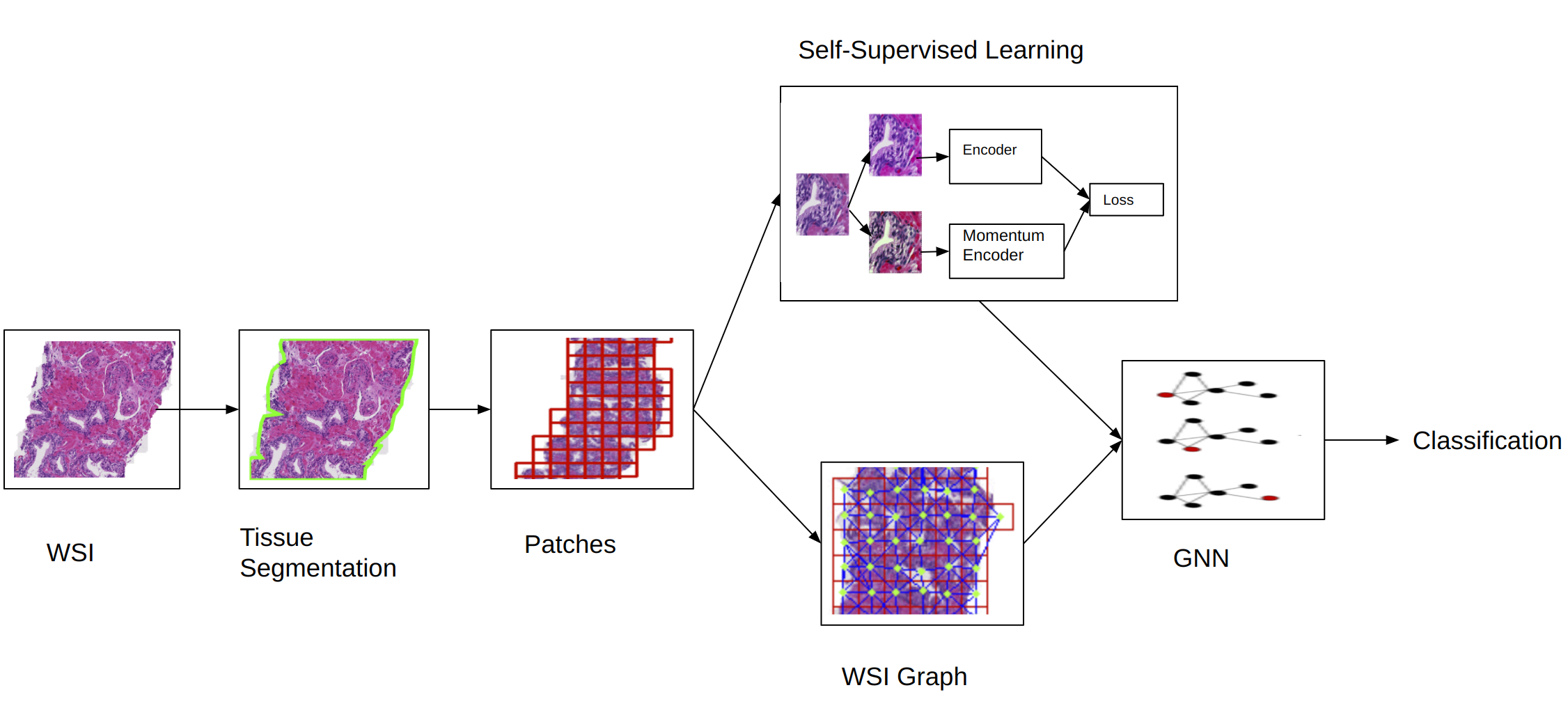}
    \caption{The proposed model's system block diagram. Given the WSI, the tissue is segmented first from the background. On the segmented tissue patches of size 256$\times$256 are generated. Each patch acts as a node for the the graph generation. Then, these patches are trained using SSL for representation learning. From SSL, feature for each patch is generated which is the node feature for our graph. The WSI graph is then trained using GCN for cancer grading. }
    \label{fig:diag}
\end{figure*}
Even though CNN is well adopted in image classification and has cutting-edge techniques for processing high-resolution pictures, they cannot be effectively trained for high-resolution WSIs. This is mostly due to the fact that a single WSI comprises more than a billion pixels at its greatest resolution, and downsampling WSIs or looking only at regions of interest may result in the loss of important information in the area surrounding the tumor that is essential for the cancer diagnosis.

The difficulties of training the WSIs can be summed up as follows:
\begin{itemize}
    \item Large dimensionality of the WSIs.
    \item Lack of annotated data.
\end{itemize}
We address these issues in this work by presenting GCN based self-supervised learning for WSIs. Self-supervised learning allows to learn the meaningful representation presented by the data without having to label or annotate the data. The tumor environment and neighborhood information are employed in training when presenting  WSIs as a graph. In summary the contribution of this work can be listed as:
\begin{itemize}
    \item Introduction of context-aware self-supervised learning on patches and graph-based learning on WSIs.
    \item Learning the features in patch levels and representation of any arbitrary size WSIs as a graph in full resolution.
\end{itemize}
\texttt{}

The remaining sections of the paper are organized as follows. The related research in this area has been covered in Section II. Section III presents our proposed approach and details about learning methods used for the experiments. Datasets used for the experiment and details about the parameters used in our training are discussed in Section IV.  The findings of the simulations are shown in section V. Discussion of the suggested work in Section VI helps to clarify the findings. Section VII serves as the paper's conclusion.

\begin{figure*}[btp]

  \centering
  \centerline{\includegraphics[width=\textwidth]{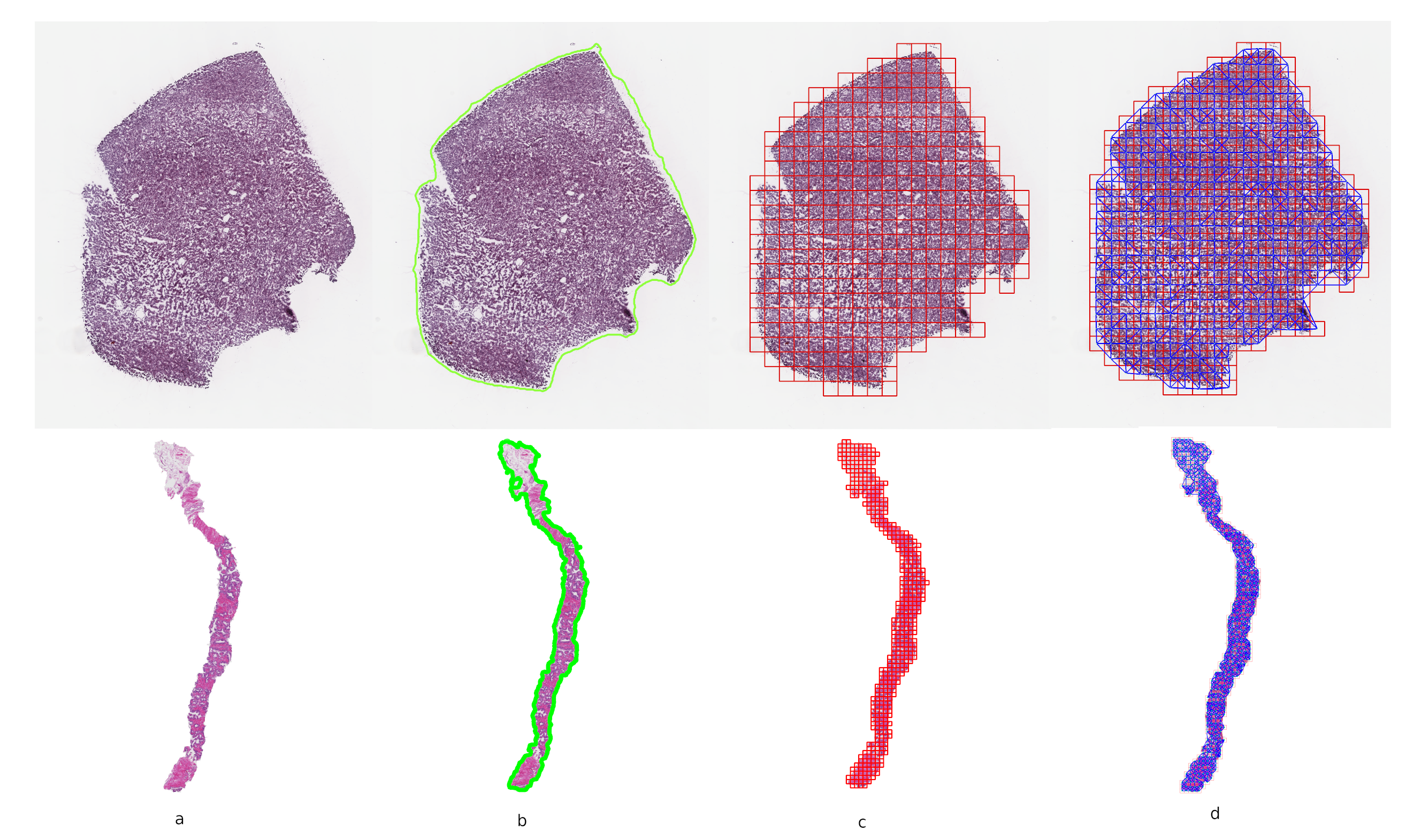}}
  \medskip
\caption{Pre-processing of WSI to generate the graph. Top part shows the graph generation in kidney resection, bottom part represents similar process in prostate biopsy. a) Whole Slide Image b) Contour separating tissue from background c) WSI divided into patches  d) WSI as a graph.}
\label{fig:preprocess}
\end{figure*}

\section{Related Work}
Recently, there has been extensive research efforts in computational pathology. In order to tackle the difficulty of training models employing WSIs, weakly supervised learning based on MIL~\cite{DIETTERICH199731} or tile-based patches have been utilized \cite{multiinstance}, \cite{nature-weaklyMIL}, \cite{Lu19}, \cite{MIL1}, \cite{mil2}. Rather than depending on fully supervised learning which requires manual annotation, the authors have proposed classifying cancer based on the WSI using MIL approach where the whole slide is divided into different tiles. The important idea is that if the WSI is negative, at least one of the tiles must also be negative and not contain the tumor. Similar to this, for the slide to be positive if the WSI is positive, the tumor must be present in at least one of the tiles. This generalizes to the so-called MIL approach.

In \cite{nature-weaklyMIL} the authors have proposed a two-stage network. They began by training the model with MIL in order to produce tile-level feature representations.  On the basis of these representations, an estimate of the final classification score was generated with the help of a recurrent neural network (RNN). The paper focuses on binary classification.  More specifically, the authors' focuses were on prostate, skin and breast cancers. Even though authors investigate the problem of prostate cancer, the model is not trained for Gleason grading. In \cite{LI2021104253}, the authors presented multi-resolution MIL, which can be trained using slide levels for prostate cancer grading. The work consists of a two-stage cancer detection model. In the initial stage, low resolution patches were examined to see if they contained any signs of malignancy. At the second stage, cancer grade classification was carried out using a higher magnification for the suspicious spots.

In \cite{Kidney} the authors look at the problem of classifying the sub-types in the kidney RCC. For the classification, they suggest a two-stage process. First the patches were extracted and a binary classification is performed to identify the cancerous and non-cancerous patch. A 3-way classification for patches was carried out using the feature extracted from the first stage in the second step. In \cite{kcancer}, a pyramidal deep learning pipeline with three CNN and patches of three different sizes were looked into to separate clear cell RCC from papillary cell RCC.    

The methods described above are looking at WSIs in patches and are solving the cancer classification task only in slide-level. Almost all the above methods use MIL. They are not context-aware as the outcome of one patch has no effect on other patches in most of the cases.  These methods rely on CNNs and are unable to capture the context between the patches. At this point, having a graph structure that is capable of capturing the relation that exists between the various nodes will come in handy. With recent advancements in graph neural networks(GNNs), they have been able to make impacts in computer visions and medical imaging application as well. In \cite{AHMEDTARISTIZABAL2022102027} authors have discussed about the various graph-based methods used in computational histopathology. A combination of MIL and graph-based learning is presented in \cite{Zhao_2020_CVPR}. The authors first selected the bag of patches and then the bag representation was presented for the graph-based MIL. A graph-based framework for the WSI is provided in \cite{rlHisto}, where the graph is built via patch selection from certain areas of WSI. This work further demonstrates that graph-based strategies perform better than MIL methods.


\section{Method}
\label{sec:method}
In Fig.~\ref{fig:diag}, the overall proposed model block diagram is presented. The block diagram demonstrates how a GCN and a self-supervised learning model are coupled for the classification of cancer. In this work, we are representing WSI, a multi-resolution image into graph data structure. The graph data to be trained into GCN requires the features, so we need to capture the information embedded in those image pixels when representing the WSI as a graph. To be more explicit, WSI is shown as a graph with each node being a smaller patch that was created by subdividing WSI. Pretrained self-supervised learning is used to encode the features within these patches. These features, along with the WSI's entire graph structure, are used to train the GCN for cancer grading.

The entire graph creation \cite{chen2021whole} procedure for the WSI is displayed in Fig.~\ref{fig:preprocess}. The following sections go over patch generation, graph generation and methods to learn from them.
\subsection{Patch generation}
\label{ssec: patch}
The dataset contains WSI with multi-scale resolution. In WSIs, the white background is useless for diagnosing cancer. The first stage in patch generation from WSI is to separate the tissue from white background. From any WSI resolution, the white background may be eliminated. The tissue is separated from the background using the segmentation network based on UNet~\cite{ronneberger2015unet} style encoder and decoder. For the purpose of training the segmentation network, the smallest WSI solution was downscaled to a 512$\times$512 image. OpenCV is utilized to create the contour dividing the tissue region from the background in the WSI using the segmentation network's output. From the WSI, patches of size 256$\times$256 are created with the greatest resolution based on the contours. These patches are used to train self-supervised learning and extract meaningful features.




\subsection{Learning from patches}
\subsubsection{Self-supervised learning}
One of the most well-known unsupervised learning methods for pre-training, often known as pretext tasks, is self-supervised learning. These pretext tasks are then adjusted for subsequent tasks \cite{He_2020_CVPR}, \cite{9156540}. The most widely used form of self-supervised learning is contrastive learning \cite{chen2020simple}. Learning the representation that optimizes the agreement between similar patches and the disagreement between distinct patches is the aim of contrastive learning. Even though the similar patches might have different views, using contrastive learning we want to learn the similar representation for each of them. These different views are generated by applying the augmentation to the original patches. 

\begin{figure}[htb]

\begin{minipage}[b]{1.0\linewidth}
  \centering
  \centerline{\includegraphics[width=8.5cm]{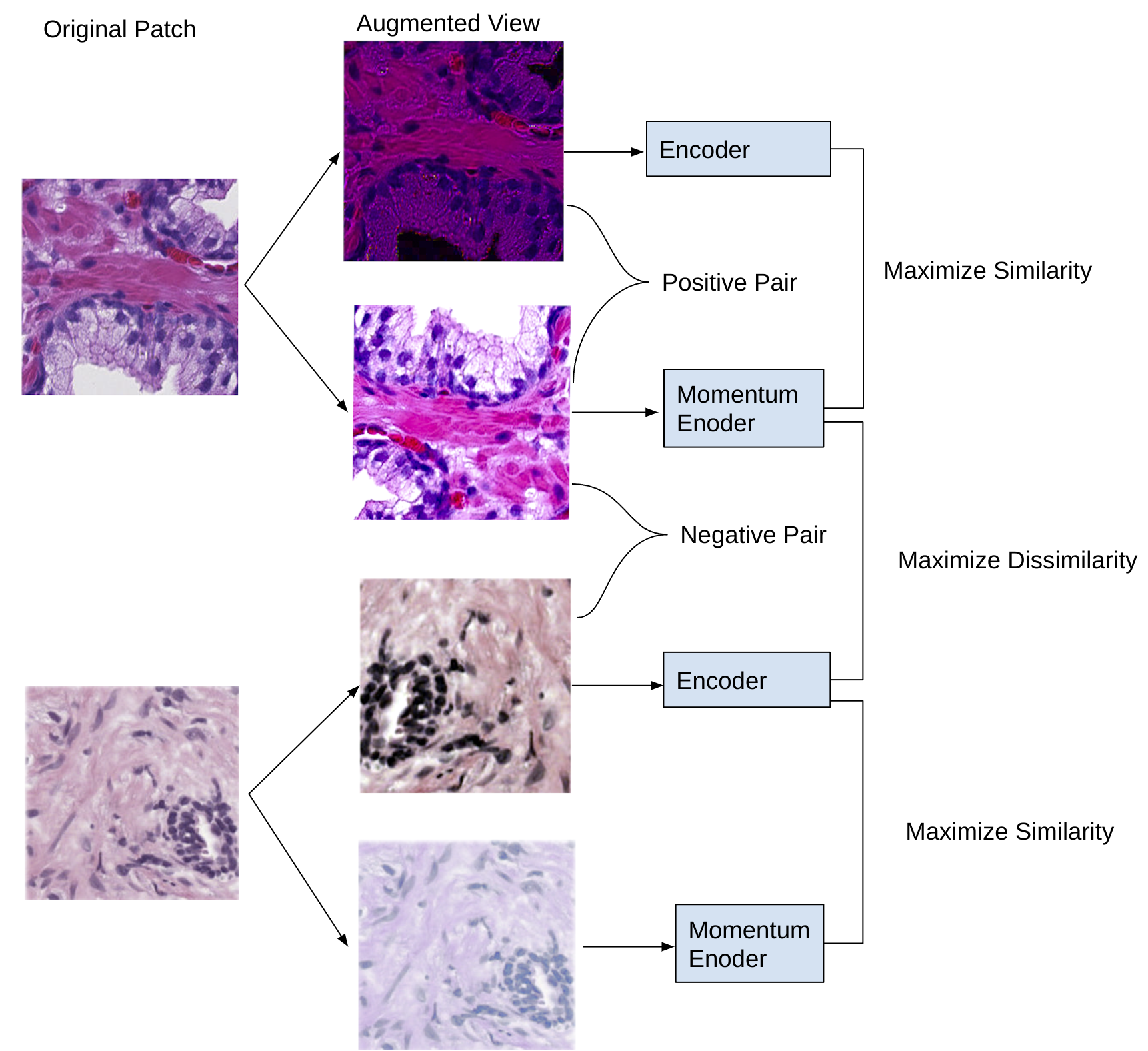}}
  \medskip
\end{minipage}
\caption{SSL framework applied for learning from the patches. The contrastive loss is used to train the SSL. The goal is to maximize the similarity between the different augmented views from the same patch, i.e., positive pairs and maximize dissimilarity between the negative pairs.}
\label{fig:SSL}
\end{figure}

Self-supervised learning techniques are best suited to learn from these patches because they are not labeled. Fig.~\ref{fig:SSL} illustrates how self-supervised learning is implemented. The original patch is applied augmentation to obtain two different views. The different views obtained from the same patch are positive pair and the the negative pairs are the ones obtained from the different patch. In this work, encoder, momentum encoder proposed in \cite{He_2020_CVPR}, \cite{chen2021mocov3} is used for the training of the self supervised network. The advantage of this method over other method such as end-to-end learning \cite{chen2020simple}, memory bank \cite{9156540} is that the batch size does not have to be large and there is no need of memory bank. Also encoder and momentum encoder share the same parameter, requiring only one model to be trained. In this method dictionary is built as a queue. The key and query can be seen as dictionary lookup.  In contrastive learning, the positive pair occurs when the key and query patches come from the same sample but the negative pair comes from a separate sample. We employ contrastive learning loss in the form of the InfoNCE provided by  ~\cite{DBLP:journals/corr/abs-1807-03748} :
\begin{equation}
    L_q = - log \frac{exp(q\cdot k^+/\tau)}{exp(q\cdot k^+/\tau)+\sum_{k^-}exp(q\cdot k^-/\tau)}
\end{equation}

where $k^+$ is the patch $q$'s positive pair and $k^-$ is the patch $q$'s negative pair, and $\tau$ is the temperature hyper-parameter.

The GCN uses features obtained from patches learnt through contrastive learning as node features. These features are learned without the need of any annotation from pathologists. Then, unlike previously used methodologies, there is no need to locate regions of interest for cancer grading or concatenate tile patches based on pixel intensity for feature extraction.


\subsection{Representing WSI as a graph}
Once patches for each of the WSI are generated and trained with the self-supervised framework then the WSI graph is generated. The WSI graph node corresponds to the centroid of each patch. The fast approximate k-nearest neighbor (K-NN) \cite{Muja09fastapproximate} method can then be used to determine the adjacency matrix for each node. The feature for each patch is extracted from self-supervised model trained on the patches of each tissue. 

\subsection{Learning on graphs}
\subsubsection{Graph Neural Network}
\begin{figure*}[htb]
  \centering
  \centerline{\includegraphics[width=\textwidth]{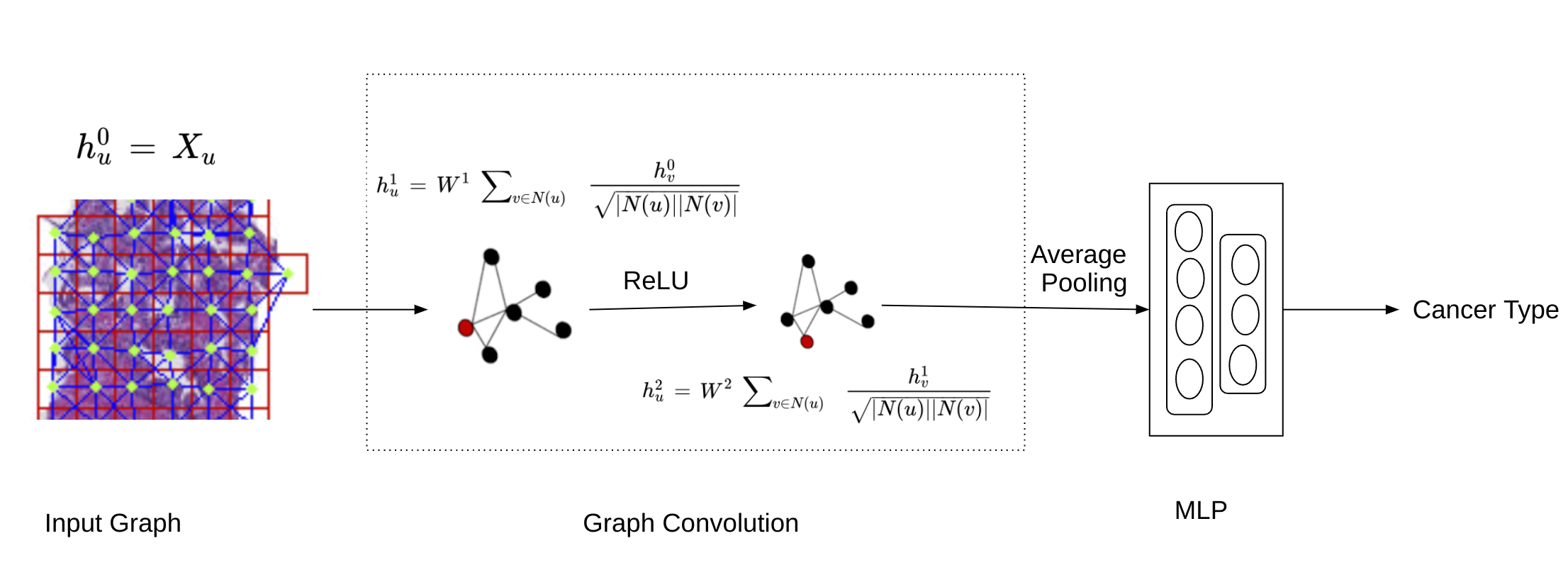}}
  \medskip
  \caption{Learning from WSI Graph using GCN.}

\label{fig:GCN}
\end{figure*}
Using graph neural learning, a meaningful representation of the graph is learned. In GNN the learning is done by passing the message between the nodes and updating it iteratively. 
The input graph $G = (V,E)$ with node features $X \in {R}^{d \times |V|}$, $d$ is the dimension of node feature vector, learns from graph neighborhood $N(u),\forall u \in V$ through message passing. The $k^{th}$ iteration message passing update is provided by\cite{William} 
\begin{equation}
\begin{aligned}
    h_u^{(k)} &= \sigma\left(W_{self}^{(k)}h_u^{(k-1)}+W_{neigh}^{(k)}\sum_{v\in N(u)}h_v^{(k-1)}+b^{(k)}\right) 
\end{aligned}
\label{eq:gnn}
\end{equation}
where the $W_{self}^{(k)}$,$W_{neigh}^{(k)}$ are trainable parameters, non-linearity is accounted by $\sigma$ and bias term is given by $b^{(k)}$. Over the course of the iterations, the embeddings $h_u$ are updated and at $k=0$, $h_u^{(0)}=X_u, \forall u\in V$. The node level equation is given by equation~\ref{eq:gnn}.  The graph level equation for message passing is as following:
\begin{equation}
        H^{(k)} = \sigma\left(AH^{(k-1)}W_{neigh}^{(k)}+H^{(k-1)}W_{self}^{(k)}\right)
        \label{eq:gnn1}
\end{equation}
where $H^{(k)}$ is a $R^{|V| \times d}$ containing feature for all nodes, graph adjacency matrix is given by $A$.
This work makes use of the GCN~\cite{NIPS2015_f9be311e} to learn from WSIs represented as graphs. Following equation gives message passing equation for GCN
\begin{equation}
    h_u^{(k)} = \sigma \left( W^{(k)} \sum_{v\in N(u)} \frac{h_v}{\sqrt{|N(u)||N(v)|}}\right)
    \label{eq:gcn}
\end{equation}
Global average pooling on all nodes is applied, then MLP head, following the implementation of the GCN layer and the use of the node feature matrix $H$ from the penultimate layer. Multi-layer perceptron is referred to as MLP. The grading of cancer is done using the cross-entropy loss function. Graph learning steps for the WSI are depicted in Fig.~\ref{fig:GCN}.

\subsubsection{GNN for WSI}
The WSIs are depicted as a graph shown in Fig.~\ref{fig:preprocess}. A tuple $G = (V,E)$ describes a graph. The nodes are represented by the set $V$, and the edges, represented by the set $E$, show how connected the nodes are to one another. In section IIA, the construction of patch is discussed. The patch is represented as the node and the the connectivity between the patches is represented as the edges.  

The WSI is a multi resolution file. We retrieved patches with a size of 256$\times$256 at the maximum resolution possible. All of the WSIs in the dataset do not have the same dimension. Because of this, there are varying numbers of nodes in the graphs due to the different numbers of patches extracted per WSI. Every node also includes a feature matrix that was created by running a patch through a previously trained self-supervised model.The acquired features are used to do learning on the graph for WSI.

The tissue and the context of the surrounding cells play a role in the cancer diagnosis. The cancerous tissue can spread over its neighbor and this information must be captured in learning. Using messages from its immediate neighborhood, a node in a GNN can learn the context of its surroundings. In our cancer grading the message is passed from its neighboring patches making our network contextually aware of its neighborhood.



\section{Implementation details}
\label{sec:pagestyle}
\subsection{Dataset}
\subsubsection{PANDA Dataset}
\begin{figure*}
    \includegraphics[width=\textwidth]{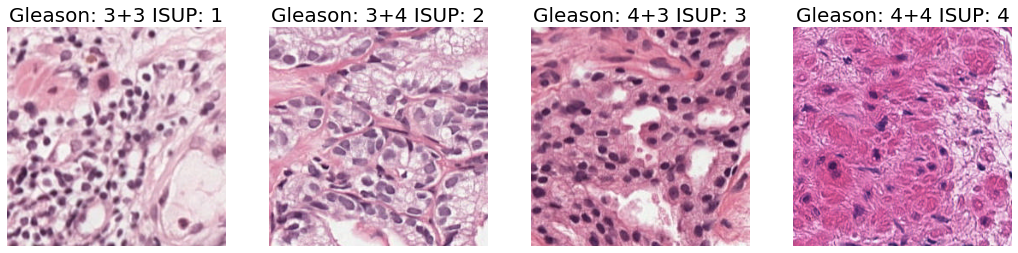}
    \caption{Patches from WSI showing Gleason and ISUP score for prostate cancer.}
    \label{fig:Gleason}
\end{figure*}

One of the datasets utilized in this work was obtained from the Kaggle PANDA challenge ~\cite{kaggle}. The WSIs of the prostate biopsies are provided by the challenge to categorize according to the Gleason score. For each biopsy two Gleason score is given by the pathologists. The most cancerous tissue is given primary score which ranges from 1-5 and secondary score which also ranges from 1-5 is assigned to other surrounding tissue. The primary and secondary score add up to give Gleason score range from 2 to 10. The International Society of Urological Pathology (ISUP) grades prostate cancer based on Gleason score. The ISUP grade along with Gleason score are displayed in Table~\ref{tab:1}. The PANDA  challenge is to assign the WSIs to one of six ISUP grades. In Fig.~\ref{fig:Gleason} the patches obtained from WSIs with Gleason score for primary and secondary scores along with ISUP grade are shown. As the glandular or white hole properties in the tissues are eliminated, as seen in Fig.~\ref{fig:Gleason}, the primary Gleason score increases. In PANDA challenge around 10500 WSIs are provided in the dataset. The resoultion scale of the WSIs in the datasets is 1, 4, 16. The highest resolution was used for the simulation in this work. The train set for our experiment contained 9500 WSIs, whereas the validation set contained the remaining ones.
\begin{table}[h]
\caption{prostate cancer grading}
\begin{center}
\begin{tabular}{|c|c|}
\hline
\textbf{Gleason Score}& \textbf{ISUP Grade} \\
\hline
6 & 1 \\
\hline
7 (3+4) & 2 \\
\hline
7 (4+3) & 3 \\
\hline
8 & 4 \\
\hline
9-10 & 5 \\
\hline
\end{tabular}
\label{tab:1}
\end{center}
\end{table}

\vspace{0.01\textwidth}
\begin{figure*}
    
    {\includegraphics[width=\textwidth]{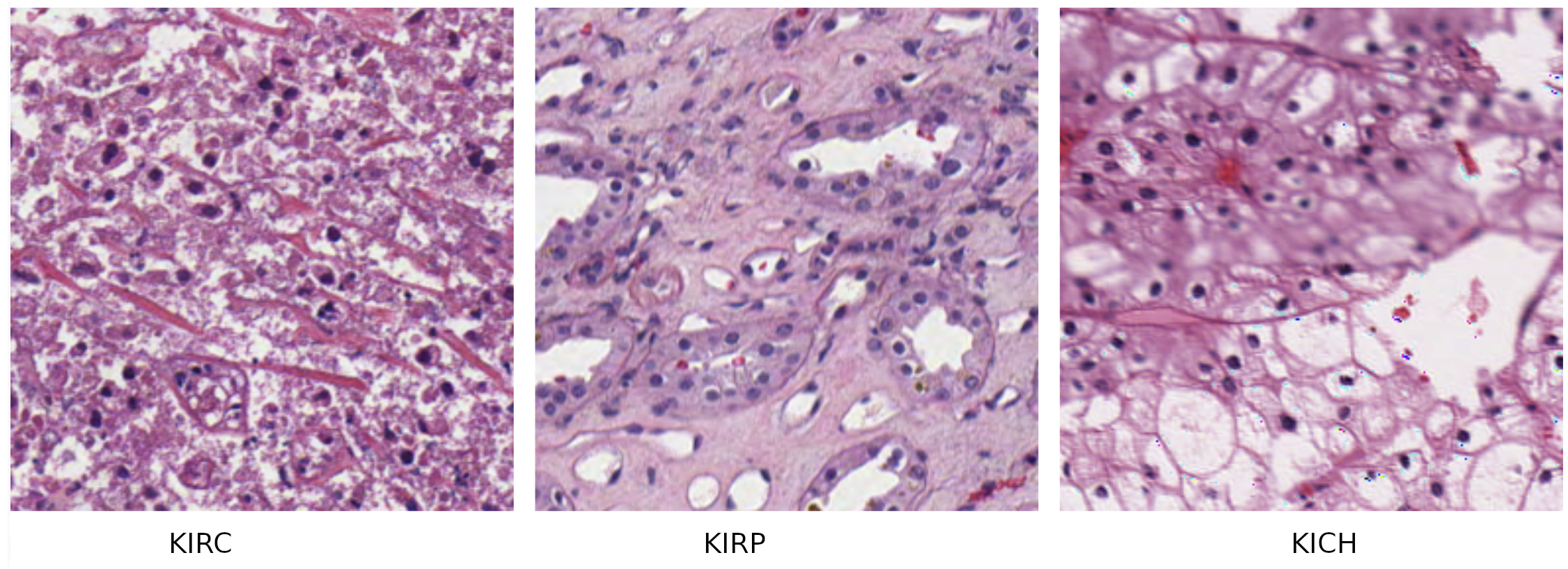}}
    \caption{Patches from WSI showing the subtype of kidney RCC.}
    \label{fig:kidney}
\end{figure*}



\subsubsection{TCGA Kidney Dataset}
The dataset for Kidney cancer was obtained from the The Cancer Genome Atlas (TCGA) portal \cite{TCGA}, \cite{TCGAkirp}, \cite{TCGAkich}. It consists of a catalog of biomedical data related to cancer. The main aim of the project is to make the data available for researchers so that various bio markers of different cancer types can be discovered. TCGA portal consists of different types of data such as clinical information, molecular analyze metadata and molecular characterization data, WSI of the tissue related to same case. In this work, we have looked into WSIs for kidney cancer. 

The TCGA dataset on kidney cancer looks into most common types of tumor in Kidney cancer i.e., RCC. The project is further grouped into most common types of RCC subtype. The TCGA consists of 3 projects for each subtype i.e., TCGA-KIRC (Kidney Renal Clear Cell Carcinoma), TCGA-KIRP (Kidney Renal Papillary Cell Carcinoma) and TCGA-KICH (Kidney Chromophobe). The most frequent malignant tumor of the kidney is renal cell carcinoma (RCC). For this work, the dataset from TCGA portal consists of 519 samples of WSIs for KIRC, 300 samples of WSIs for KIRP, and 121 WSIs sample for KICH. The train set consists of the 85\% of the data and rest is included in validation set. In Fig.~\ref{fig:kidney} the patches for the subtype of the kidney RCC is shown.

\subsection{Training the self-supervised model}
The first part of of our experiment required training the model for the self-supervised learning. Before training the self-supervised part patches were extracted from the WSI and made ready for the self-supervised training. The patches were extracted after training a small subset around 10\% of the dataset for the segmentation of tissue to separate tissue from white background using UNet network. The UNet network with EfficientNet as the backbone was trained for 20 epochs. 

The first part of the experiment involved training self-supervised model for 75 epochs. The model was trained using Adam optimizer, the weight decay was 1 $\times$ $10^{-6}$, and the learning rate was 3 $\times$ $10^{-3}$.  The learning rate was modified using the cosine scheduler.
The batch size was set as 256. During training 0.2 was used as the value of parameter $\tau$. The mocoV3 \cite{chen2021mocov3} framework was used to train the self-supervised network with ResNet50 as backbone. The contrastive learning requires different views of the same patch which were obtained using the data augmentation while training. Random Gaussian Blur, random contrast adjustment, and random horizontal and vertical flip are among data augmentation methods used during training.

\subsection{Training the graph network}
In the second part of the experiment, trained self-supervised model was used to extract the feature for each patch and create the graphs for the WSI. After the graphs were constructed for the whole dataset,  GCN was trained. For GCN Adam optimizer was used to train the model for 30 epochs. The learning rate was modified using cosine scheduler during the training with initial learning rate 1 $\times 10^{-4}$, weight decay of $10^{-6}$. The model was trained with batch size of 1.

\subsection{Evaluation Metrics}
In this work we make use of weighted Kappa ($\kappa$) \cite{kappa} score to measure the performance of our model.  The agreement between both the actual and projected outcome is measured by the quadratic weighted kappa score. 
An N $\times$ N matrix $O$ is formed such that $O_{i,j}$, $i$ is actual and $j$ is predicted outcome. 
 With a N$\times$N weight matrix as $ w_{i,j}=\frac{(i-j)^2}{N-1}$, $\kappa$ is calculated as
 \begin{equation}
     \kappa = 1 - \frac {\sum_{i,j}w_{i,j}O_{i,j}}{\sum_{i,j}w_{i,j}E_{i,j}}
     \label{eq:kappa}
\end{equation}
Where, in $O_{i,j}$, $i$ is actual and $j$ is predicted outcome.

\section{Results}
\label{sec:typestyle}
We validated our model with two different datasets. First, the model is validated using the WSIs from the biopsies of the prostate cancer and then the simulations are run on WSIs from the resection of the kidney. The $\kappa$ score was used to assess the model's performance. 

For the first test, PANDA dataset was used to train the model. Malignant slides were to be graded from 1 to 5, whereas non-cancerous slides were to be predicted a score of 0 for each WSI.

\begin{figure*}[htb]
  \centering
  \centerline{\includegraphics[width=\textwidth]{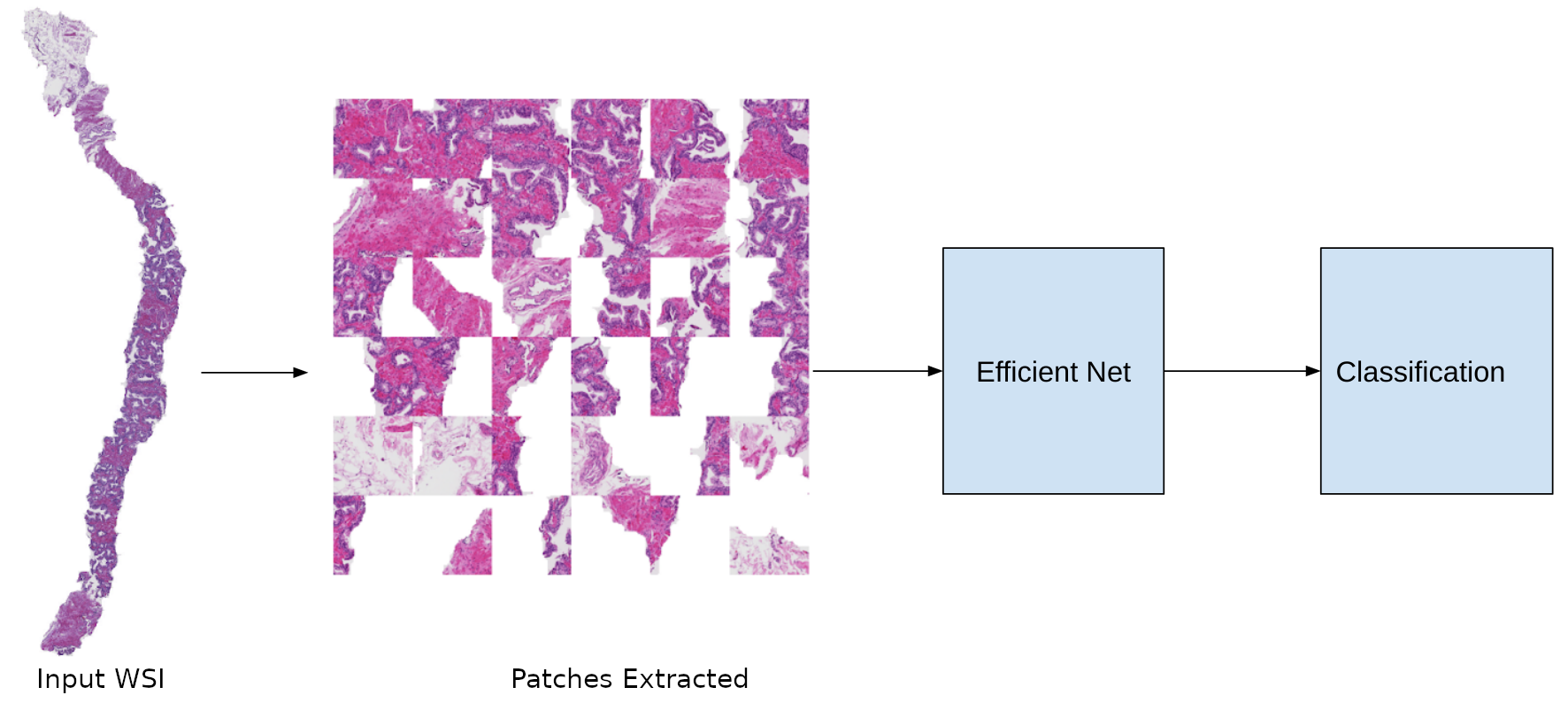}}
  \medskip
\caption{Baseline Model. WSI is converted into tiles of patches using Blue Ratio Selection. Once patches are concatenated, it is passed into EfficientNet for the classification.}
\label{fig:MIL}
\end{figure*}

Before using the proposed approach, we trained the model using only a traditional CNN, ResNet50~\cite{resnet}. Concatenated tile pooling was used to train the WSIs. The ResNet50 is trained using the concatenated 36 tiles that are chosen from each WSI. On validation datase a kappa score of 0.764 was obtained for this model. Next we tested the dataset with MIL based approach using EfficientNetB0 \cite{effnet} as CNN network for training. Kappa score increased to 0.79 using this method. In Fig.~\ref{fig:MIL} we can see the model for this implementation. From any WSI, 36 patches are selected and bagged together and this bag is passed as a single image for training. These patches are selected using the blue ratio(Br) selection given by equation~\ref{eq:BR}. The R,G,B in (6) represent the Red, Green and Blue channels in the original image. The blue ratio is widely used in selecting the useful part in the WSI \cite{br1}.

\begin{equation}
    Br=  \frac{100 \times B}{1+R+G}\times\frac{256}{1+R+G+B}
    \label{eq:BR}
\end{equation}

The transfer learning based model using EfficientNet was further evaluated using ensemble method. The model was trained using both EfficientNetB0 and EfficientNetB4\cite{effnet}. These models were then ensembled together and the new kappa score was improved to 0.87. Attention-based MIL based on the model in ~\cite{deepattn}  was used to compare with the proposed model. The kappa score for attention-based MIL with this dataset was0.883.
The performance of the model when the features of the graph were extracted with self-supervised model versus transfer learning through ResNet50 have been further compared. 
Initial tests on the proposed graph-based technique were conducted without the use of any self-supervised learning. ResNet50 was used to extract the features from the patches for GCN. Each patch was passed through ResNet trained by ImageNet \cite{deng2009imagenet} and the features were extracted. These features were then used as nodes in the GCN training of the proposed method. The GCN model with features for nodes extracted from the ResNet achieved a kappa score of 0.785.  

Then, the features for patches were extracted using the model trained in self-supervised learning. For the proposed model we extracted the feature size of 248 and 2048 from the pre-trained self-supervised model. Those features were extracted from the different layers in the encoder of the self-supervised model. 

The method suggested in this work was assessed using four-fold cross validation. Since the two graphs for each WSIs represent different node feature sizes (248 and 2048 features for each node in the graph), two GCN networks were trained. The kappa score for the proposed model with nodes having 248 features was 0.871. The GCN model with 2048 features obtained a kappa score of the 0.891. Next, the goal was to evaluate the performance when these two models were ensembled. The kappa value of 0.899 was reached by ensembling the two models. When compared to the simple tile-based technique, this presents a significant improvement. The results of numerical experiments are summarised in Table~\ref{tab:2}.
\begin{table}[h]
\caption{Comparison of Kappa Score for Prostate Cancer dataset. The proposed method is compared with existing MIL-based and attention-based approaches. The proposed GCN-based algorithm has been tested with features extracted from ResNet and the proposed self-supervised approach.}
\begin{center}
\begin{tabular}{|c|c|}
\hline
\textbf{Method} & \textbf{PANDA Dataset} \\
\hline 
ResNet50\cite{resnet}  & 0.764   \\
\hline 
MIL with Efficinet Net \cite{2018} & 0.79   \\
\hline
MIL with Ensembled Efficinet Net & 0.87 \\
\hline
Attention based MIL~\cite{deepattn}  & 0.883 \\
\hline
GCN with features from ResNet   & 0.785 \\
\hline
GCN with 248 Features for each Node (Proposed) & 0.871 \\
\hline
GCN with 2048 Features for each Node (Proposed) & 0.891 \\
\hline
GCN with Ensembled Model(Proposed)  & \textbf{0.899} \\
\hline
\end{tabular}
\label{tab:2}
\end{center}
\end{table}

The algorithm  was further tested using a dataset from TCGA. The WSI from the TCGA were kidney resections which consisted of larger resolution scale compared to the prostate biopsies. The WSIs were converted to graphs with 20$\times$ resolution. All of the overall algorithm steps were repeated for this dataset as well. 

Using kappa score, the performance of different models for Kidney cancer dataset, is compared in Table~\ref{tab:3}. Similar to prostate dataset, initially the dataset was modeled using ResNet50. The kappa score of 0.758 was achieved. This dataset was further evaluated using MIL with EfficientNet. The blue ratio selection of patches for MIL was used for this dataset and a kappa score of 0.783 was obtained. Similar to the first experiment, multiple MIL-based methods were ensembled to achieve the kappa score of 0.868. The kappa score for this data set when tested on attention based MIL was 0.896. Then, the proposed graph-based methods were evaluated on this dataset. For the GCN model with ResNet50 used as feature extractor a kappa score of 0.792 was obtained. The kappa score was improved to 0.887 when self-supervised model was used as feature extractor with 248 features for each node in GCN. The score improved to 0.924 with 2048 features for each node. These two models were ensembled to obtain a score of 0.939. Overall results demonstrate that the proposed self-supervised methods with GCN show a better performance compared to other existing approaches.

\begin{table}[h]
\caption{Comparison of Kappa Score for different method applied to Kidney Cancer dataset}
\begin{center}
\begin{tabular}{|c|c|}
\hline
\textbf{Method}&\textbf{Kidney Dataset} \\
\hline 
ResNet50 \cite{resnet}  & 0.758  \\
\hline 
MIL with Efficinet Net \cite{2018} & 0.783  \\
\hline
MIL with Ensembled Efficinet Net &0.868 \\
\hline
Attention based MIL~\cite{deepattn} & 0.896 \\
\hline
GCN with features from ResNet  & 0.792\\
\hline
GCN with 248 Features for each Node (Proposed) & 0.887  \\
\hline
GCN with 2048 Features for each Node (Proposed) &0.924 \\
\hline
GCN with Ensembled Model(Proposed) & \textbf{0.939}\\
\hline
\end{tabular}
\label{tab:3}
\end{center}
\end{table}

\section{Discussion}
In this work, GNN-based method that uses self-supervised model to extract features of the nodes has been developed. The validity of the proposed model was checked by the utilization of two distinct cancer datasets.  To be more explicit, the performance of the suggested model is assessed by utilizing the WSIs that were obtained from the biopsies of the prostate as well as the resections of the kidney. Two different models of the GCN were developed, one of which is based on the size of the features for each node. The first GCN model was trained using the WSI graph, which had a feature size of 248 for each node in the graph. The feature size of the node in the graph for the second GCN network that was employed in the training of the WSI graph was 2048. After each model had been trained independently, they were brought together to form an ensemble.  The proposed model obtained a kappa score of 0.899 for the grading of prostate cancer and 0.939 for the grading of sub-type RCC in the diagnosis of kidney cancer.

The absence of data annotation is a significant obstacle for those working in the medical area, in particular with WSI, due to the fact that it is a laborious task for the pathologist to provide annotation in detail. We tackled this issue by extracting features from patches without any annotation using self-supervised learning. Additionally, the dimensionality of WSI presents another challenge. The entire WSI cannot be trained because of its enormous dimension. The training of the WSI using GCN can be accomplished in its entirety by first subdividing the WSI into smaller patches and then representing each patch as a node in the graph.

We also compare our model with other traditional models that are commonly used for the training of the WSI. First, we evaluated an MIL-based method that uses a certain number of patches selected based on certain rules such as blue ratio selection. Applying this approach has disadvantages as it cannot capture all parts of the WSI and only a fixed number of patches are selected to train using CNN. The grading of the cancer might be different if a single patch of cancerous tissue is missed. This problem is solved by using the graph data structure to represent the WSI. Another advantage of using GCN is that it takes into account the context of the tissues in WSI. Experiments show that the proposed method outperforms MIL-based approaches in terms of performance.

Using self-supervised learning to extract features for the patches leads to a considerable improvement in the kappa score. According to the findings from this study, the GCN that was trained using features obtained through self-supervised learning performed significantly better than the GCN that was trained using features derived from the ResNet50. This provides more evidence that, when using self-supervised learning, there is no requirement to manually increase the amount of annotated data. This framework is well-motivated for learning from WSIs and is broad enough to address the issue of insufficient annotation for other types of medical images.

\section{Conclusion}
\label{sec:majhead}
The work that is being presented here aimed to achieve the goal of being able to represent any WSI of arbitrary size and learn from them without having to make substantial use of the annotations. The use of a self-supervised learning method in conjunction with a GNN allows for the achievement of these goals.  

In this work, we suggested a two-stage model for the classification of cancer using WSIs. The approach would consist of a self-supervised model and a GCN. In order to depict WSIs as a graph, it is divided up into smaller patches. These patches are first trained in self-supervised models. As node features for the WSI graph, the self-supervised model's learnt features for the patches were applied. The reconstructed graph along with features is trained with GCN. The context of each cell and its surroundings were then included into this graph by GCN training in order to grade and diagnose cancer. Without further pathologist annotation, our method enables learning of the features from the patches. Additionally, the usage of GCN makes it possible to learn WSIs in their entirety.

\section*{Acknowledgment}

The training of models in this work were made possible by Raj High Performance at Marquette University. 

The results published here are in part based upon data generated by the TCGA Research Network:  https://www.cancer.gov/tcga.


\ifCLASSOPTIONcaptionsoff
  \newpage
\fi



%
\bibliographystyle{IEEEtran}
\bibliography{journal}



%








\end{document}